\documentclass[11pt]{article}
\usepackage{amsmath,amsthm,latexsym,amssymb,amsfonts,parskip,epsfig}

\usepackage{palatino}
\usepackage[mathcal]{euler}

\newcommand{\R}{\mathbb{R}}                  
\newcommand{\Z}{\mathbb{Z}}                  
\newcommand{\N}{\mathbb{N}}                  

\newcommand{\aver}[1]{\langle #1 \rangle}

\title{Toward explaining black
hole entropy quantization in loop quantum
gravity}
\author{Hanno Sahlmann, Spinoza Institute/ITP, Utrecht University}
\date{{\small Preprint ITP-UU-07/46, SPIN-07/34}}

\begin{document}
\maketitle
\begin{abstract}
In a remarkable numerical analysis of the
spectrum of states for a spherically symmetric
black hole in loop quantum gravity, Corichi,
Diaz-Polo and Fernandez-Borja found that the
entropy of the black hole horizon increases in
what resembles discrete steps as a function of area.
In the
present article we reformulate the combinatorial
problem of counting horizon states in terms of
paths through a certain space. This formulation
sheds some light on the origins of this step-like behavior
of the entropy. In particular, using a few
extra assumptions we arrive at a formula that
reproduces the observed step-length to a few
tenths of a percent accuracy.

However, in our reformulation the periodicity
ultimately arises as a property of some
complicated process, the properties of which, in
turn, depend on the properties of the area
spectrum in loop quantum gravity in a rather
opaque way. Thus, in some sense, a deep
explanation of the observed periodicity is still
lacking.
\end{abstract}
\section{Introduction}
Recently, a large computerized analysis of the
spectrum of states for a spherically symmetric
black hole in loop quantum gravity was carried
out by Corichi, Diaz-Polo and Fernandez-Borja
\cite{Corichi:2006wn,Corichi:2006bs,Corichi:2007zz,DiazPolo:2007gr}.
The analysis focused in particular on the entropy
for the black hole. The theory
\cite{Rovelli:1996dv,Ashtekar:1997yu,Ashtekar:2000eq,
Domagala:2004jt,Meissner:2004ju,Ghosh:2006ph,Ghosh:2004wq}
predicts the area-dependence of the entropy as
\begin{equation}
\label{eq_basic} S(A)=\frac{\gamma_c}{\gamma}
\frac{A}{4l_P^2} -\frac{1}{2}
\ln\left(\frac{A}{l_P^2}\right) +O(A^0).
\end{equation}
$\gamma$ is the Barbero-Immirzi-parameter, and
$\gamma_c$ a numerical constant\footnote{$\gamma_c$
depends on the precise definition of
which states are counted as surface states. There
are two possibilities, leading to two slightly
different values for $\gamma_c$: Just counting
the $m$- and $b$- labels (ex.
\cite{Ashtekar:2000eq,Domagala:2004jt}) leads to
$\gamma_c\approx 0.237$, whereas also counting the
$j$-labels (ex. \cite{Ghosh:2004wq}) leads to
$\gamma_c\approx 0.274$.}. Fixing $\gamma$ to be
equal to $\gamma_c$ insures consistency with the
Bekenstein-Hawking entropy law.
The numerical analysis
\cite{Corichi:2006wn,Corichi:2006bs,Corichi:2007zz,DiazPolo:2007gr}
was confined to relatively modest black hole
sizes, up to a few hundred Planck-lengths squared
in terms of the black hole area. Still, it
required the calculation of up to $10^{58}$
states. The work beautifully confirmed the
leading order behavior linear in $A$ and it was
even powerful enough to confirm the next to
leading order contribution. But what is more, it
discovered a completely unexpected periodic
behavior of the entropy: There is a very
distinctive periodic signal superimposed onto the
linear growth \eqref{eq_basic}. Its period was
found to be proportional to $\gamma$,
\begin{equation}
\label{eq_period} \Delta A=\gamma\chi l_P^2,
\end{equation}
where $\chi$ was estimated in \cite{Corichi:2006bs} to be
\begin{equation}
\label{eq_cdf} \chi\approx \chi_\text{CDF} = 8.80.
\end{equation}
This signal leads to a stair-case like behavior for the
entropy as a function of the area and was hence
called \textit{entropy quantization} in
\cite{Corichi:2006wn}.

The observation of this period is remarkable in
several ways. For one, \eqref{eq_period} can be
read as a sort of \textit{effective} equidistant
quantization of area \cite{Corichi:2006wn}. Thus
although the area spectrum in loop quantum
gravity is \textit{not} equidistant but rather
consists of sums of terms of the form
\begin{equation}
\label{eq_specc} A_j=8\pi\gamma l_P^2
\sqrt{j(j+1)},\qquad j\in \N/2.
\end{equation}
\eqref{eq_period} does provide a point of contact
with the ideas of Bekenstein (see for example
\cite{Bekenstein:1995ju,Bekenstein:1998aw}) on
area quantization. This point gets even more
pronounced when one takes into account that, as
first observed in \cite{Corichi:2006wn}, the
value $\chi_{CDF}$ is very close to
$8\ln(3)\approx 8.788898$. Based on a heuristic
quantization of the black hole and the
Bekenstein-Hawking-relation for entropy and area, an
equidistant area spectrum had been conjectured
\cite{Bekenstein:1998aw}, with
\begin{equation}
\label{eq_heuristic}
\Delta A= 4 \ln(k) l_P^2
\end{equation}
and $k$ an integer. Later,
$k=3$ was suggested by Hod \cite{Hod:1998vk}
using a correspondence to the frequencies of quasinormal
modes of the black hole in a certain limit. Dreyer \cite{Dreyer:2002vy}
observed that in loop quantum gravity one could
reinterpret \eqref{eq_heuristic} as the smallest nonzero
area eigenvalue, making a certain choice for the parameter
$\gamma$ that seemed natural at that time, but
it was later realized that that $\gamma$ would not lead to
the Bekenstein-Hawking-relation for entropy and area. Given all this,
it is intriguing that an equidistant spectrum with
$\Delta A$ involving $\ln(3)$ seems to reappear here.
We should not fail to mention however, that there is
an additional factor of $2\gamma$ in \eqref{eq_period}
as compared to \eqref{eq_heuristic}, so ultimately
it is not clear wether there is a deep correspondence
with the ideas of Bekenstein and Hod.
For more comments on this see \cite{Corichi:2006wn},
for some considerations of physical consequences
we refer to \cite{DiazPolo:2007gr}.

Secondly the phenomenon seems to be robust
against the way the the split between surface and
bulk degrees of freedom is done. As indicated in
\eqref{eq_period}, $\Delta A$ just depends on
$\gamma$, and $\chi$ (and not on $\gamma_c$), and
is hence independent of the way the states were
counted.\footnote{It is only \textit{a
posteriori} that one might decide to adjust the
freely specifiable parameter $\gamma$ to take the
value $\gamma_c$ and hence make \eqref{eq_basic}
consistent with the Bekenstein-Hawking entropy
formula.}

Finally, there does not seem to be a
straightforward way to explain this phenomenon
from the theory. After all, as \eqref{eq_specc}
shows, the area spectrum is quite complicated,
non-equidistant, and does not, in any obvious
way, determine the constant $\chi$. We should note
that the $A_j$ of \eqref{eq_specc} do become
approximately equidistant for large $j$ (for
details see \cite{Sahlmann:2007jt}), so it might
at first seem that this gives an explanation
for the observed periodicity.
However, the spacing in \eqref{eq_specc} becomes
close to multiples of $4\pi \gamma l_P^2
\approx 12.56\,\gamma l_P^2$.
Comparing this with \eqref{eq_cdf}, it is clear
that this can not furnish a simple explanation
for the value of $\chi$.

With the present paper we aim to contribute to an
explanation of this phenomenon of entropy
quantization. We use two main ideas. The first is
to reformulate the combinatorial problem of
enumerating the physical states for the black
hole horizon in terms of paths built from a set
of elementary steps. The second idea is to use a
statistical description of the set of paths, very
similar to a random walk.

These ideas, together with some assumptions on
the numerical distribution of steps will enable
us to calculate an approximation to the period
$\Delta A$ that reproduces the observed
step-length to a few tenths of a percent
accuracy. $\Delta A$ arises as some sort of `resonance'
in the area spectrum \eqref{eq_specc}. While we
think that this is a nice result, it is not
a complete analysis and explanation of the
phenomenon. To start with, since we have to make
some assumption and approximations, our result
for $\Delta A$ is not exact, and its uncertainty
is hard to determine. Thus we are unable to
confirm or rule out that $\chi=8\ln(3)$.
Furthermore our analysis will not determine
wether the phenomenon will persist for
arbitrarily large areas. We will
discuss these points in more detail, below.

After the present work was finished, we became aware
of results \cite{Ansari:2006vg} by Ansari. His is a very
nice analysis of the spectrum of the \textit{full} area
operator, breaking it down into an infinite set
of equidistant sub-spectra. For the black hole horizon,
only a subset of that spectrum is relevant, due to the
horizon boundary conditions, and the exclusion of edges
that are a submanifold of the horizon, as well edges that
pierce the horizon from within the black hole.
Ansari's methods break down for this ``reduced''
area operator. Still it is not inconceivable that his results
are connected to the entropy quantization,
and thus may
represent another way of looking at the problem
tackled in the present work.

Finally we would like to bring to the reader's
attention that there is independent very interesting
work on the way \cite{CDF} which, using somewhat
similar methods, will shed more light on the phenomenon
of entropy quantization.

The paper is organized as follows: In the next
section we review the combinatorial problem of
enumerating horizon states of the black hole and
formulate it in terms of steps and paths. Section
\ref{se_deltaa} is concerned with statistical
considerations and with a computation of $\chi$.
We finish with a discussion of the results and
future perspectives in Section \ref{se_disc}.
\section{Counting states by counting paths}
In the present article, we will not review any of
the physics behind the description of an isolated
horizon in loop quantum gravity. We refer the
interested reader to
\cite{Ashtekar:1997yu,Ashtekar:2000eq}. Here it
will suffice to spell out the combinatorial
problem to which finding physical states of the
black hole horizon is ultimately reduced. The
quantity of greatest interest to us is $N(I)$,
the number of horizon states that are eigenstates
of horizon area with eigenvalue in the interval
$I$. As we have said in the introduction, there
are two ways to count states, depending on where
one draws the line between bulk and boundary
degrees of freedom. Both lead to qualitatively
similar results for $N(I)$. Here we will consider
only one of these ways, the one that was laid out
in \cite{Ashtekar:2000eq} in detail. In
\cite{Domagala:2004jt} this problem was revisited
and given a very simple formulation. It was shown
that $N(I)$ is the number of ordered sequences
$(m_i)_i$, $m_i\in \Z_*/2$ such that
\begin{equation}
\label{eq_bla} \sum_i m_i=0 \qquad
\text{and}\qquad 8\pi\gamma l_P^2
\sum_i\sqrt{|m_i|(|m_i|+1)}\in I.
\end{equation}
The connection to the quantum geometry of the
horizon is that sequences $(m_i)_i$ with these
properties are labels of physical states of the
horizon. Let us simplify even further and get rid
of all the units, by defining $n(a)\doteq
N(8\pi\gamma l_P^2 a)$. It will also be useful to
introduce the shorthand
\begin{equation*}
a(m) \doteq \sqrt{|m|(|m|+1)},\qquad m\in \Z/2.
\end{equation*}
Finally we take from
\cite{Domagala:2004jt,Meissner:2004ju} the idea
that the counting problem can be simplified by
implementing the two conditions of \eqref{eq_bla}
in separate steps. We define
\begin{equation*}
n(a,j)\doteq\left\lvert \left\{(m_1,m_2,\ldots
),\, m_i\in \Z_*/2\, : \, \sum_i m_i=j, \sum_i
a(m_i)= a\right\}\right\rvert.
\end{equation*}

It will be instructive to reinterpret the state
labels as paths. To this end, introduce the space
$\mathcal{S}\doteq\R_+\times\Z/2$. Let us call a
sequence of points $(p_i)_i$ in this space a
\textit{path}, and the differences $p_{i+1}-p_i$
the \textit{steps} of the path. Now let us call a
path $(p_i)_i$ \textit{allowed} if
\begin{itemize}
\item it starts at 0, i.e. $p_1=(0,0)$, and
\item the steps $p_{i+1}-p_i$ are of the form $v(m_i)\doteq (a(m_i),m_i)$ for some
$m_i\in\Z/2$.
\end{itemize}
Obviously, then, we can associate to a state
labeled by $(m_1,\ldots m_n)$ the allowed path
$(0, v(m_1),v(m_1)+v(m_2),\ldots)$.

What we are really interested in is the number
$n(I)$ of states with the area in an interval
$I$. In the language of paths this is given as
\begin{equation*}
n(I)= \text{the number of allowed path with
endpoint in } I \times \{0\}.
\end{equation*}
Given any number $R\geq 0$ there is a finite
number of allowed paths that end within
$[0,R]\times\Z/2\subset\mathcal S$. It is easy to
enumerate them, and we have written a little
\textit{Mathematica} routine that does this for
us.
\begin{figure}
\centerline{\epsfig{file=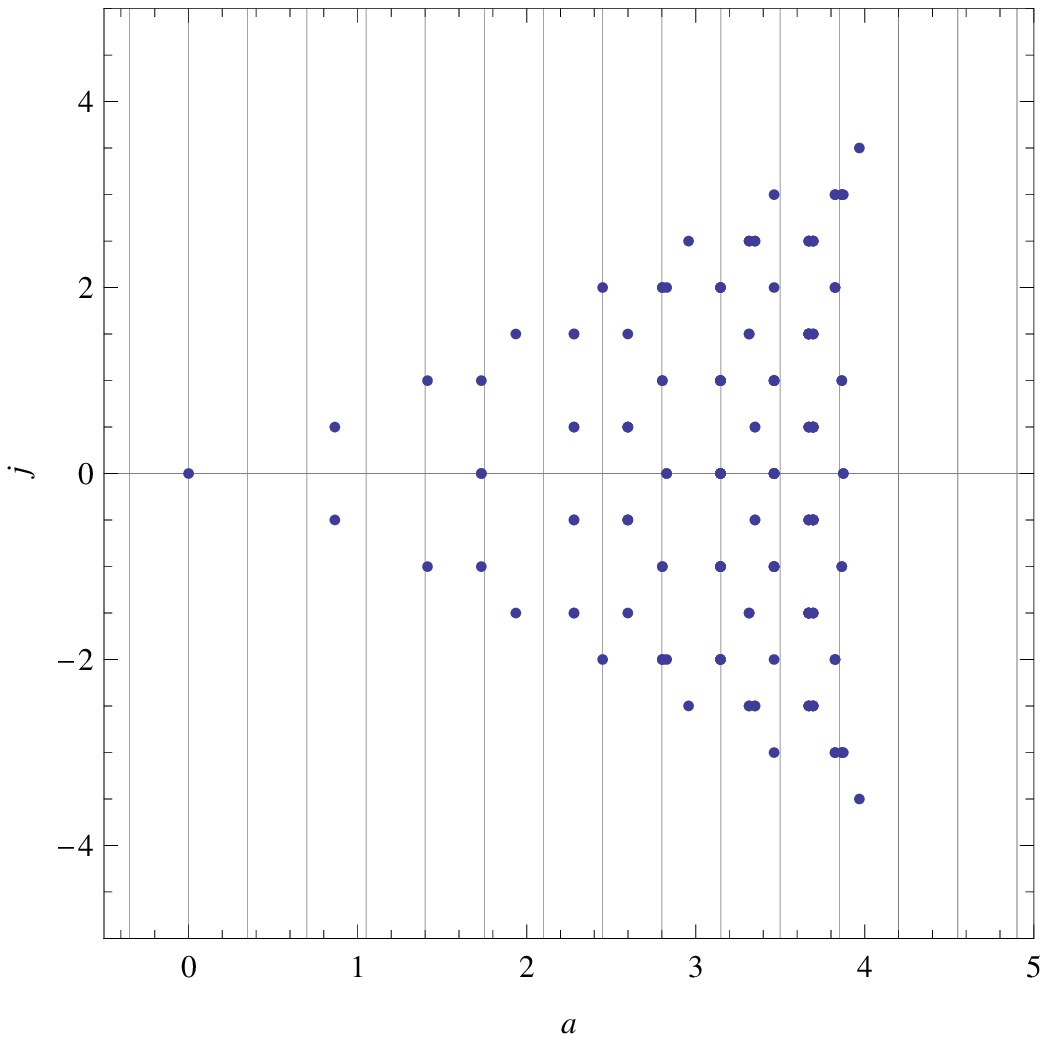,
scale=0.6}$\quad$ \epsfig{file=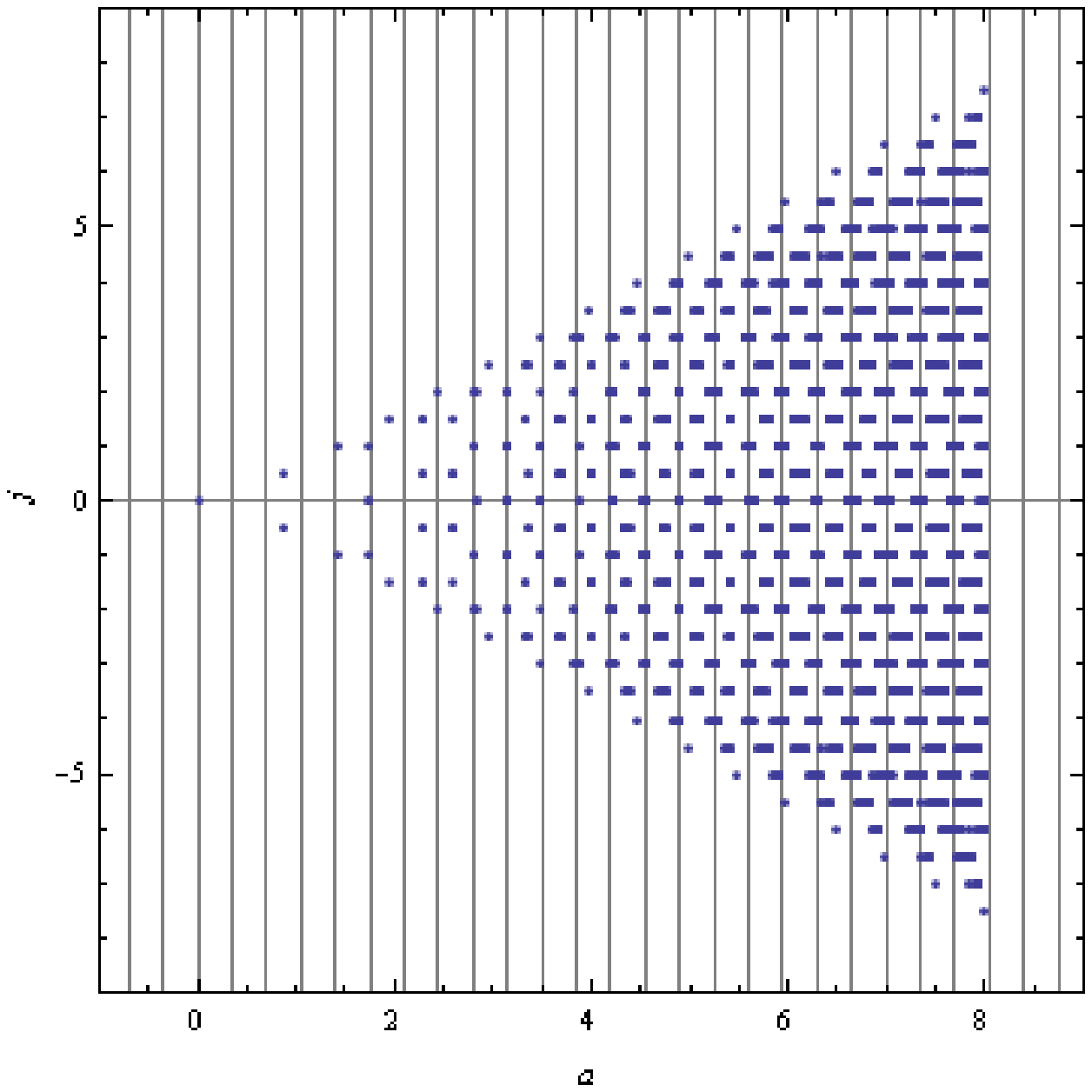,
scale=0.5}} \caption{\label{fi_tree} Paths that
end below $R=4$ (left) and below $R=8$ (right).}
\end{figure}
In Figure \ref{fi_tree} we show the results, for
$R=4$ and $R=8$, by plotting, for each path, its
endpoint $(a,j)$ as a dot in the diagram. In
effect these diagrams contain all information about
the functions $n(I)$ and $n(a,j)$. The most
obvious feature of the results is the striking
regularity that they exhibit. We should mention that
many of the points plotted in Figure
\ref{fi_tree} lie on top of each other. For $R=8$
there is for example a total of 76619 points to
plot. So the regularity in the concentration of
points becomes even more striking if one plots
the \textit{density} of endpoints of paths.
\begin{figure}
\centerline{\epsfig{file=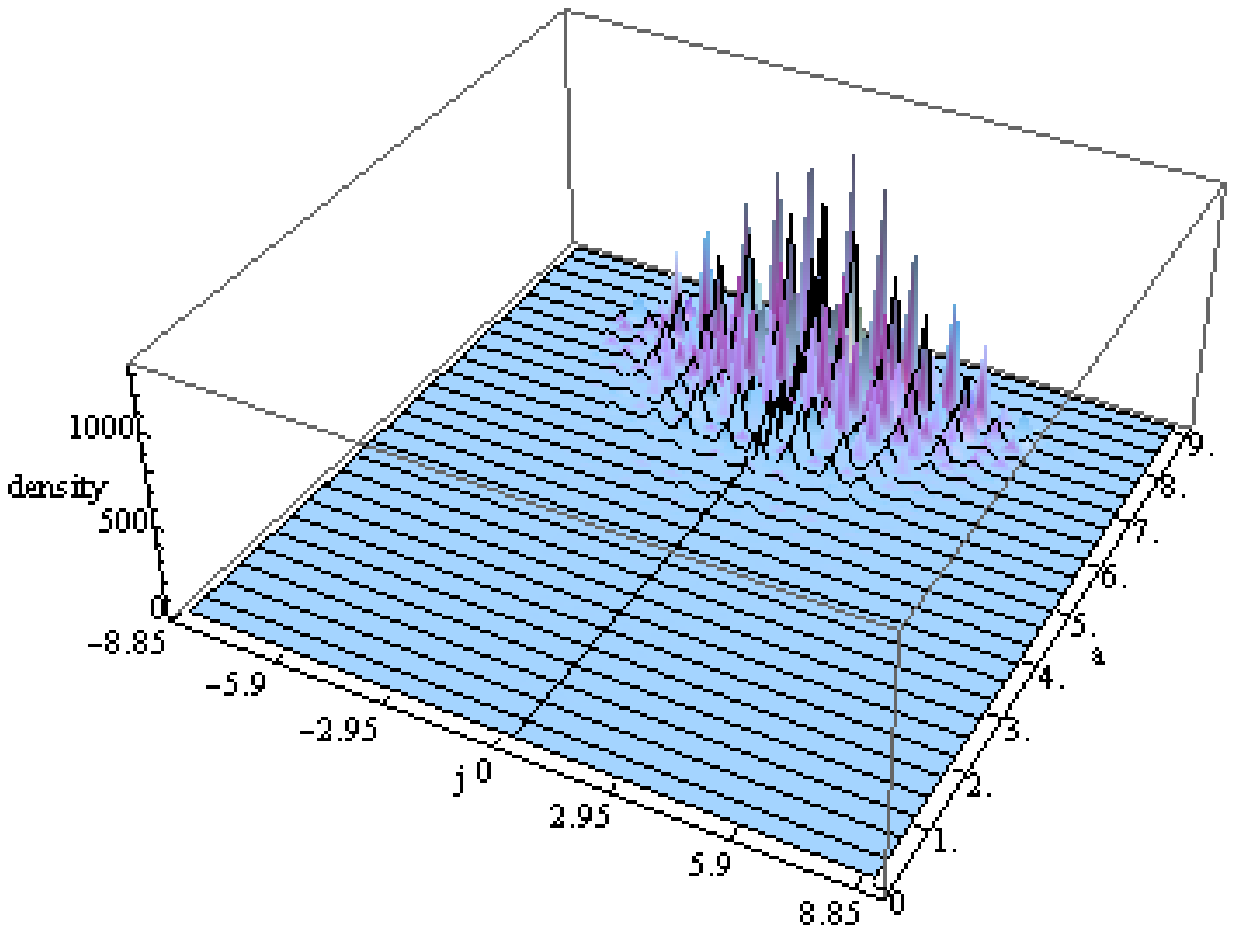,
scale=0.5}$\quad$ \epsfig{file=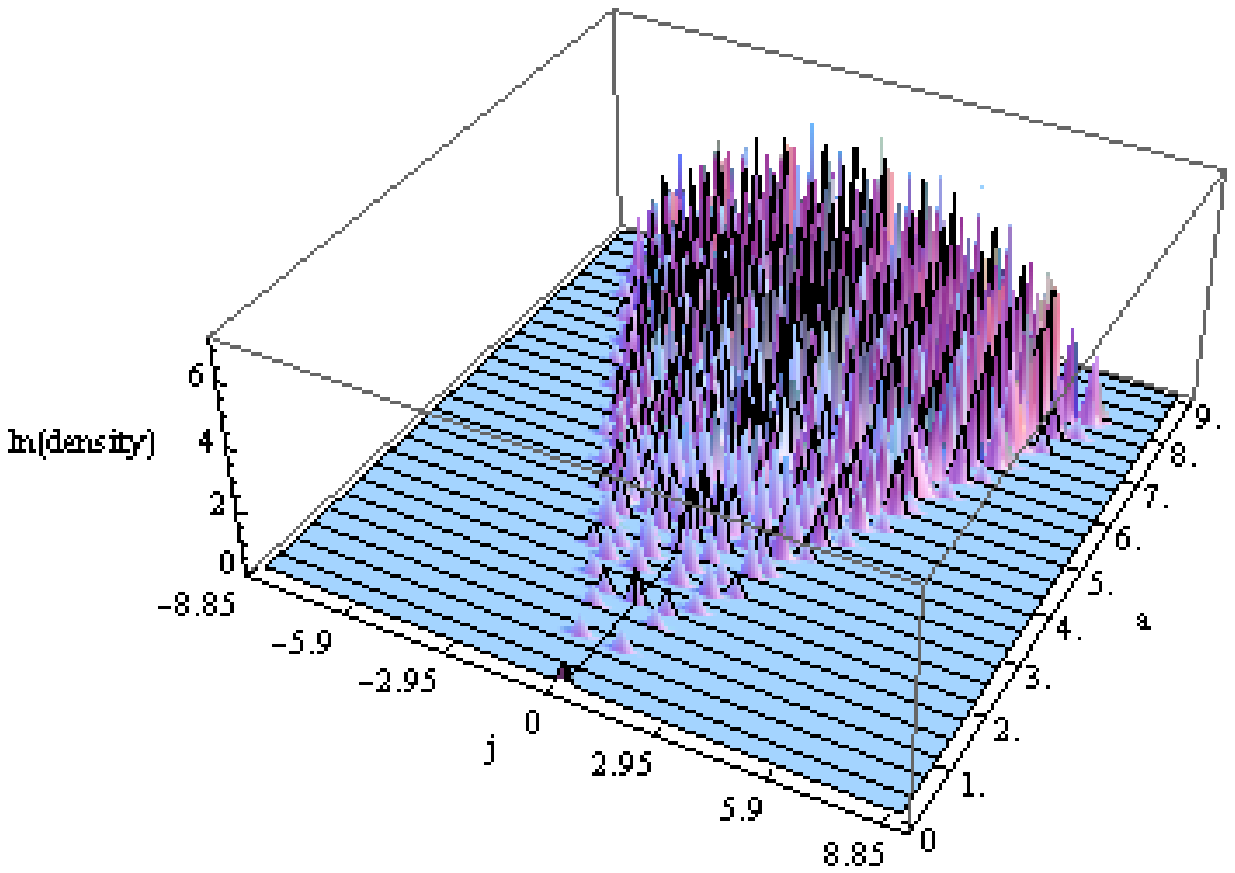,
scale=0.5}} \caption{\label{fi_density} Density
(left) and logarithm of density (right) of paths
that end below $R=8$ (in fiducial units).}
\end{figure}
We have done so in Figure \ref{fi_density} for
$R=8$.

Now an important question is the following: Is
the period we see in these figures the one that
was observed in
\cite{Corichi:2006wn,Corichi:2006bs,Corichi:2007zz,DiazPolo:2007gr}?
The answer is yes: The $a=$ \textit{const}. lines drawn in the
Figures \ref{fi_tree} and \ref{fi_density} are
regularly spaced at intervals
\begin{equation*}
\Delta a= 0.35.
\end{equation*}
Their correspondence to the period
exhibited in the data is clear. Moreover,
physical states are associated to the points on
the line $\R_+\times \{0\}$ in the diagrams. Thus
the periodicity on that line corresponds directly
to the periodicity seen in
\cite{Corichi:2006wn,Corichi:2006bs,Corichi:2007zz,DiazPolo:2007gr}.
\begin{figure}
\centerline{\epsfig{file=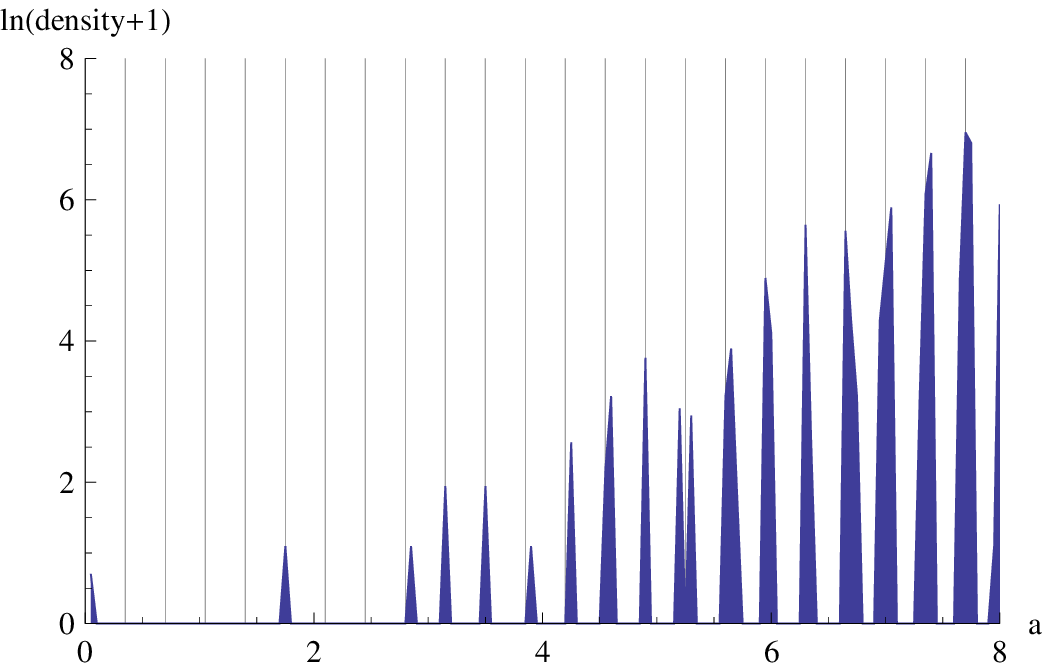,
scale=0.6}$\quad$ \epsfig{file=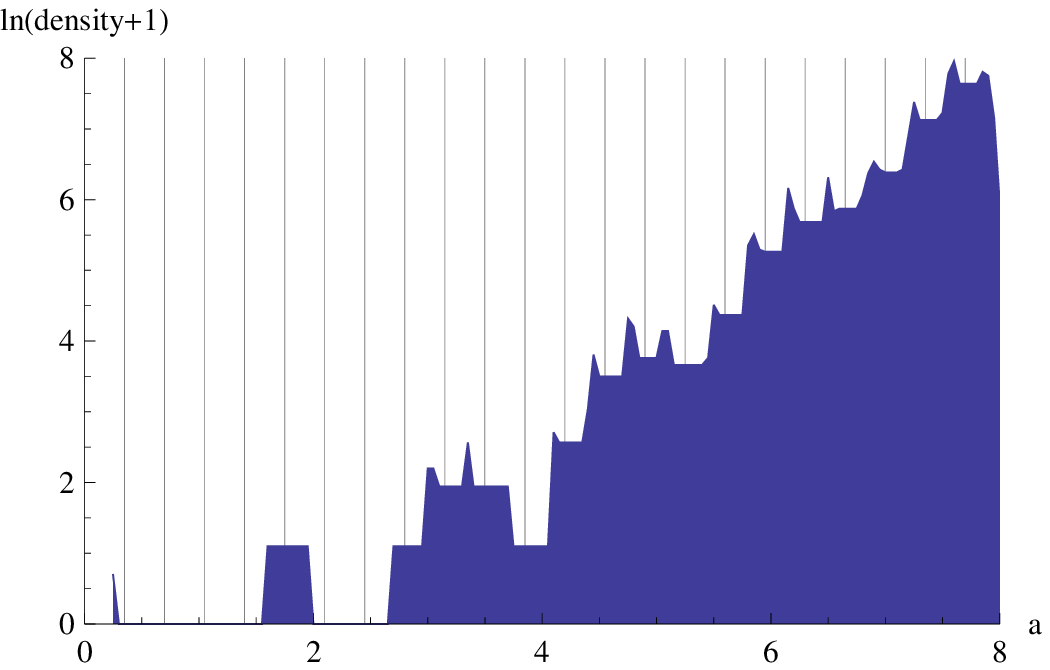,
scale=0.6}} \caption{\label{fi_stair} Logarithm
of density of physical states, in a small
interval around $a$ (left) and in an interval
adapted to the periodicity (right).}
\end{figure}
We have also plotted density on this line and
logarithm of density in a sliding interval
$[a-\Delta a/2, a+\Delta a/2]$ (Figure
\ref{fi_stair}. The latter starts to show the
staircase of
\cite{Corichi:2006wn,Corichi:2006bs,Corichi:2007zz,DiazPolo:2007gr}
near $a=8$.

It must be said that the amount of data that we have
assembled is much smaller than that handled
in the much more sophisticated analysis
\cite{Corichi:2006wn,Corichi:2006bs,Corichi:2007zz,DiazPolo:2007gr}.
Based on our data alone it would be premature to
conclude that a periodicity in the entropy is
present. We are however confident that our plots
show the onset of the pattern that
\cite{Corichi:2006wn,Corichi:2006bs,Corichi:2007zz,DiazPolo:2007gr}
has demonstrated much more clearly and to much
higher values of area.

What we have done so far certainly does not
amount to an explanation of the periodicity. We
merely reformulated the problem of enumerating
surface states of the black hole into one
concerning paths in the space $\mathcal{S}$. We
then observed that the periodicity found in
\cite{Corichi:2006wn,Corichi:2006bs,Corichi:2007zz,DiazPolo:2007gr}
does apparently not just govern paths corresponding to
physical states (the line $j=0$ in the figures)
but a larger class of paths. In the next section
we will show that using the image of steps and paths can
be very helpful in the analysis of the pattern.
\section{$\Delta A$ and the statistics of the steps}
\label{se_deltaa} Now that we have exhibited the
pattern in our reformulation through paths and
steps, let us return to its explanation. The
reformulation can shed new light on the issue as
follows: Imagine for a moment that all the steps
that would be allowed in allowed paths were just
integer multiples of one basic step. Then
regularity in a diagram like Figure \ref{fi_tree} would
obviously result. This is not the case for the
system at hand: The allowed steps $v(m)=(a(m),m)$
are \textit{not} integer multiples of one
another. Moreover, even if the steps were
approximately multiples of one basic step (as
could be argued is the case for the area spectrum
\cite{Sahlmann:2007jt}), stringing together many
steps would in general lead arbitrarily far away from points
in a regular pattern. So at first sight the
consideration of a situation with just integer
multiples of a basic step does not seem to lead
anywhere in terms of explaining the pattern
observed here and in
\cite{Corichi:2006wn,Corichi:2006bs,Corichi:2007zz,DiazPolo:2007gr}.
We should however keep in mind that what we want
to explain is not a completely rigid phenomenon.
It is not so that there are \textit{no} states
that fall outside the pattern observed in Figure
\ref{fi_tree}, it is rather that the majority
clusters around some evenly spaced points.
Moreover the observed pattern is really the
result of thousands (or in the case of
\cite{Corichi:2006wn,Corichi:2006bs,Corichi:2007zz,DiazPolo:2007gr},
more like of $10^{40}$) points, which in our
language are each obtained by taking many steps.
So what we should be looking for is not a formula
that describes all the details of the spectrum,
but something that explains why it is
\textit{statistically} likely for a point to lie
in one of the clusters.

Coming back to the steps and paths, our idea is
as follows:  If we can demonstrate that the steps
are multiples of a single step \textit{on
average} in a suitable sense and moreover that
the variance of the steps around this average is
very small, then we can at least explain that a
pattern formed at low areas will reproduce
itself for some time. Let us try to make this argument more
precise.

Let us assume that there is a well defined
probability distribution $p(m)$ for the
occurrence of a step $v(m)$ in a path
corresponding to a physical state. Let us
furthermore assume that we can treat the
individual steps in a given path corresponding to
a physical state as independently distributed and
with the distribution $p(m)$, to a good
approximation. Then let us write
\begin{equation}
\label{eq_ansatz} a(m)=I(m)\Delta a + \epsilon(m)
\end{equation}
where $I(m)$ shall be an integer and
$\epsilon(m)<\Delta a$. Now we consider a path
with $n$ steps $v(m_i)$ that starts somewhere on
the lattice $\Delta a \Z\times Z/2$. The distance
of the endpoint of this path to the corresponding
lattice point is
\begin{equation*}
\delta(n,\{v(m_i)\}) = \sum_{i=1}^n
\epsilon(m_i).
\end{equation*}
Now we look at this quantity under the
probability distribution. Because of our
assumptions we can use the central limit theorem
to approximate
\begin{equation}
\label{eq_central} \aver{\delta(n)}\approx
n\aver{\epsilon(m)},\qquad
\aver{\delta(n)^2-\aver{\delta(n)}^2}\approx
n\aver{\epsilon(m)^2-\aver{\epsilon(m)}^2},
\end{equation}
where the averages on the left of these
approximate equalities are expectation values in
the ensemble of physical paths with $n$ steps,
whereas the averages on the right are in the
ensemble of steps,
\begin{equation*}
\aver{f}\doteq \sum_{m\in \Z_*/2} f(m)p(m)
\end{equation*}
for $f$ a function on $\Z_*/2$.
\eqref{eq_central} shows that if we can choose
$\Delta a$ in \eqref{eq_ansatz} such that
$\aver{\epsilon(m)}=0$ then we can expect the
path to remain near the lattice as long as
\begin{equation*}
\sqrt{n\aver{\epsilon(m)^2}} < \Delta a,
\qquad \text{or} \qquad n < \frac{(\Delta
a)^2}{\aver{\epsilon(m)^2}}.
\end{equation*}
Let us apply this reasoning to the problem at
hand. What we need is (a) information about the
statistics of the steps involved, and (b) we will
have to make an ansatz for the function $I(m)$ of
\eqref{eq_ansatz}.

As for (a), luckily we have at least some
information on the statistics of the horizon
states. In \cite{Domagala:2004jt} it was shown
that upon picking a path at random out of all
physical paths ending below some $a_0$, the probability
to find $v(m)$ as the first step is
\begin{equation*}
p(m)\approx\exp \left(-2\pi \gamma_M
\sqrt{|m|(|m|+1)} \right)
\end{equation*}
with $\gamma_M\approx 0.2375$ the numerical
constant that makes the sum of the $p(m)$ over
$m$ equal 1, and the approximation good as long
as $a(m)$ is small compared with $a_0$. What we
would rather like to know is a different
probability, namely that of finding $v(m)$ as
first step in a path \textit{among all the paths
of $n$ steps} and ending below $a_0$. Now, as long
as $n$ is large, (but not as large as $a_0$), the
dependence of this probability on $n$ should be
rather weak, and thus we assume that it is
proportional to the $p(m)$ above. Thus we will
work with an ensemble of steps with the above
probability distribution, i.e. we will define the
average for a function $f$ on $\Z_*/2$ by
\begin{equation*}
\aver{f}\doteq \sum_{m\in \Z_*/2}
f(m)\exp\left(-2\pi \gamma_M \sqrt{|m|(|m|+1)}
\right).
\end{equation*}

\begin{figure}
\centerline{\epsfig{file=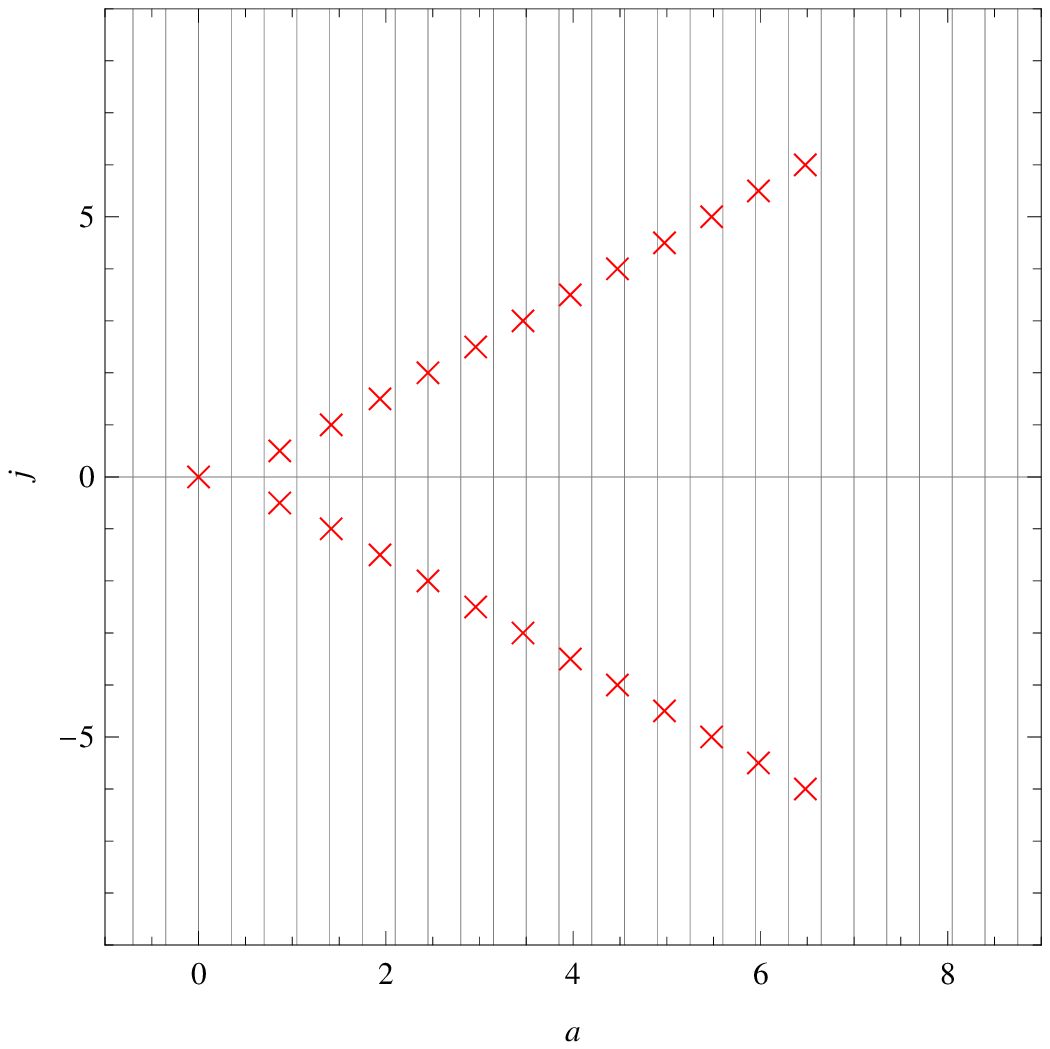,
scale=0.6}$\quad$\epsfig{file=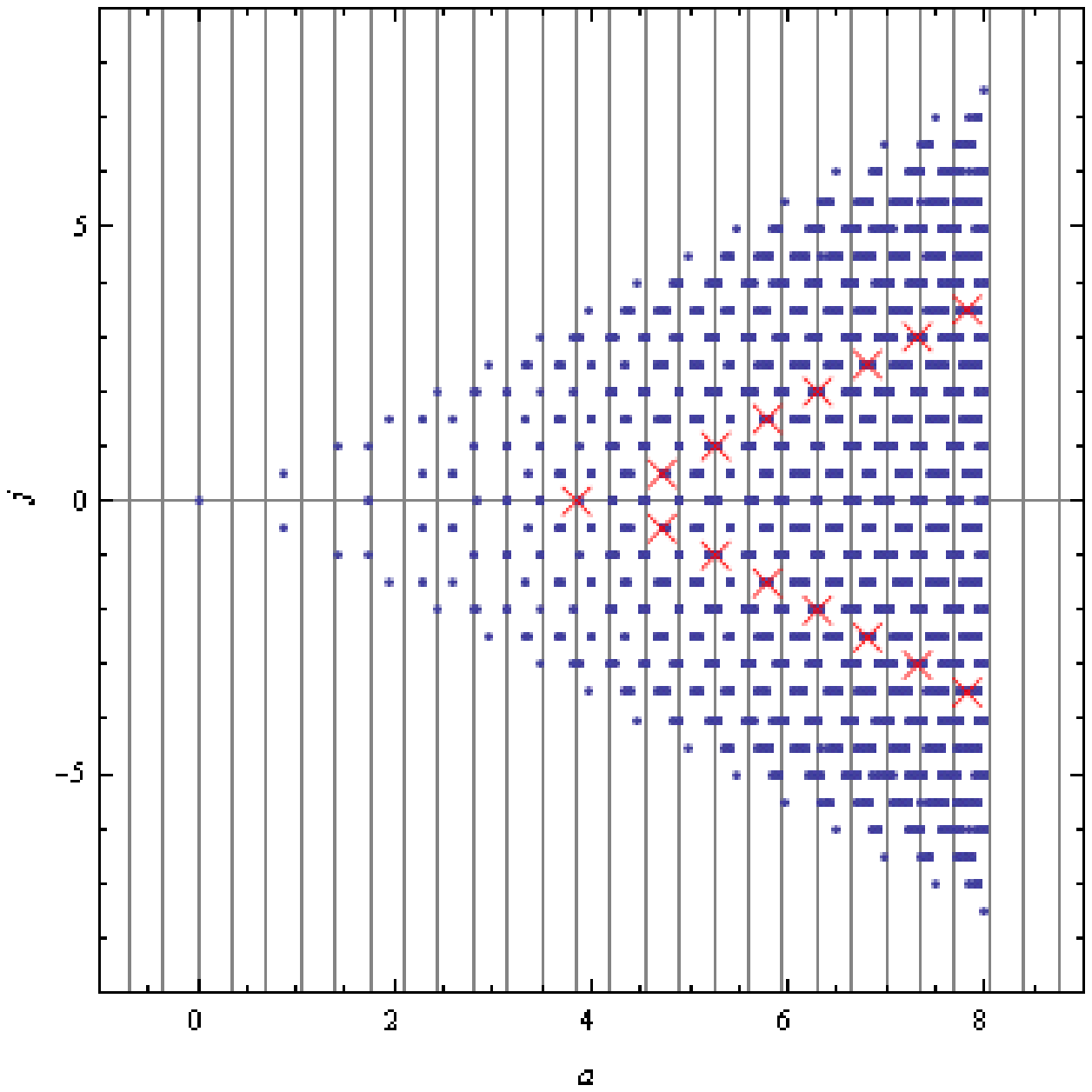,
scale=0.5}} \caption{\label{fi_teewith} The
possible steps starting at 0 (left) and a set of these steps
superimposed onto the spectrum of paths (right)}
\end{figure}
Now we turn to part (b): We have to make an
Ansatz for the function $I$ of \eqref{eq_ansatz}.
To that end, we inspect Figure \ref{fi_teewith}
which shows the basic steps, and how they fit
into the pattern of the allowed paths. One sees
very clearly that the regularity in the pattern
of the paths and of the steps are related. More
precisely $a(m+1/2)-a(m)\approx 3 \Delta a/2$.
Moreover there is a shift of one unit,
independent of $m$. Altogether, we will write
\begin{equation}
\label{eq_ansatz2} a(m)=\left(\frac{3}{2} \cdot 2
m + 1\right)\Delta a +\epsilon(m).
\end{equation}
which in turn defines the quantities
$\epsilon(m)$, once $\Delta a$ is fixed.

Now we can proceed as outlined above. We want to
determine $\Delta a$, and we want to do it in
such a way that $\aver{\epsilon(m)}$ is zero,
otherwise any pattern would be washed out. Taking
averages of \eqref{eq_ansatz2} and using the
condition $\aver{\epsilon(m)}=0$ indeed
determines $\Delta a$:
\begin{equation*}
\Delta a = \frac{\aver{a(m)}}{3\aver{m}+1}
\end{equation*}
This in turn fixes
\begin{equation*}
\epsilon(m)=a(m)-(3m+1)\frac{\aver{a(m)}}{3\aver{m}+1}.
\end{equation*}
Numerical evaluation of these formula can be done
very easily on a computer. We find
\begin{equation}
\Delta a \approx 0.34952,\qquad
\aver{\epsilon(m)^2}\approx 0.00019156
\end{equation}
Let us put these numbers in perspective and into
context. First of all, we find that the standard
deviation for the $\epsilon(m)$ is very small:
\begin{equation*}
\frac{\Delta a}{\sqrt{\aver{\epsilon(m)^2}}}
\approx 25
\end{equation*}
That means that only after a number $n$ of steps
of the order of  $625(=25^2)$ do we expect to
deviate from the pattern substantially, as we
have argued before. This means that at the very
least our results are significant for the black
holes of small area as considered here and in
\cite{Corichi:2006wn,Corichi:2006bs,Corichi:2007zz,DiazPolo:2007gr}.
Secondly, our value for $\Delta a$ compares
nicely with the one obtained in
\cite{Corichi:2006wn,Corichi:2006bs,Corichi:2007zz,DiazPolo:2007gr}.
Let us illustrate this in terms of the parameter $\chi$.
It is related to our $\Delta a$ by $\chi=8 \pi
\Delta a$, which compares to the result
$\chi_\text{CDF}$ of
\cite{Corichi:2006wn,Corichi:2006bs,Corichi:2007zz,DiazPolo:2007gr}
as follows:
\begin{equation*}
\chi\approx 8.7843, \qquad \chi_\text{CDF}\approx
8.80\qquad
\frac{\chi_\text{CDF}-\chi}{\chi_\text{CDF}}\approx
0.00129,
\end{equation*}
so we agree with the quoted reference to within a
fraction of a percent. What is more, we seem to
be even closer to the conjectured value
$8\ln(3)$:
\begin{equation*}
8\ln(3) \approx 8.7889, \qquad
\frac{8\ln(3)-\chi}{\chi} \approx 0.00053.
\end{equation*}
We will discuss these findings further in the
next section.
\section{Discussion and outlook}
\label{se_disc}
What we have done in the present paper is to
give an explanation of the phenomenon of entropy
quantization in loop quantum gravity by means of
a formulation using paths, steps, and their
statistics. The periodicity in the spectrum of
horizon states arises as some sort of
\textit{resonance} \eqref{eq_ansatz2} in the area
spectrum.\footnote{We should stress that although
we talk about a \textit{resonance}, and our model for the
black hole involves some sort of random walk, these
are not \textit{physical} processes. They have nothing
to do with the dynamics of the black hole.}
That \eqref{eq_ansatz2} works so well
has to do with the fact that the area spectrum is
nearly equidistant, $\sqrt{j(j+1)}\approx j+1/2$.
The precise relation \eqref{eq_ansatz2} that gives
small $\aver{\epsilon(m)^2}$ (and hence the value of $\Delta A$)
does however
depend on the details of the area spectrum as well
as on the quantum boundary conditions (i.e.\ the $j=0$ constraint). Thus it
is intimately related to
properties of the area quantization of loop
quantum gravity, however in a rather opaque way.
\begin{figure}
\centerline{\epsfig{file=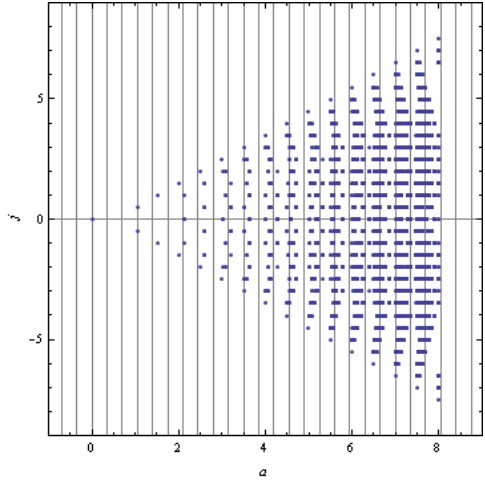, scale=0.5}}
\caption{\label{fi_badtree} The set of paths
ending below $R=8$, with a distorted area
spectrum $a'(m)$}
\end{figure}
A nice way to confirm this is to redo the
calculation of allowed path, however using a
slightly distorted area spectrum $a'(m)\doteq
a(m)+1/(10 m)$. One can see that the details
of the result (Figure \ref{fi_badtree}) change
quite drastically as compared to the undistorted
spectrum. One does however still recognize a lot
of regularity in the result. Our interpretation
is that because the area spectrum is still
approximately equidistant, one again sees
regularities emerge, whereas the precise pattern
has changed because the resonance condition
to achieve small $\aver{\epsilon(m)^2}$
is now different.
%

Our explanation seems to be quite successful
quantitatively, as we recover the results of
\cite{Corichi:2006wn,Corichi:2006bs,Corichi:2007zz,DiazPolo:2007gr} for $\chi$.
However here already one problem of our approach
becomes apparent: Several approximations go into
the determination of our value for $\chi$, and we
have little idea how accurate the result actually
is.

As for other aspects of the phenomenon, some can
be explained by our approach, while others remain
mysterious. In particular, we want to remark the
following:

(1) The phenomenon occurs for both ways to count
\cite{Corichi:2006wn,Corichi:2006bs,Corichi:2007zz,DiazPolo:2007gr}
but here we have considered only one. We note
however that the other way of counting states
(i.e. the inclusion of $j$-labels) can be seen as
a \textit{refinement} of the counting we have
done here. To be more precise, each path
corresponding to a physical state that we have
counted here corresponds to \textit{one or more} physical
states as counted in the other scheme. Thus the
pattern that we have observed is bound to appear
also in the other scheme, possibly modulated
further by some other effects that come from the
details of the counting of the $j$-labels. Thus
it seems to us that the explanation of the
phenomenon given here also applies to the other
counting scheme.

(2) It was observed
\cite{Corichi:2006wn,Corichi:2006bs,Corichi:2007zz,DiazPolo:2007gr}
that the phenomenon goes away when not
implementing the condition that $\sum_i m_i =0$.
From our Figures \ref{fi_tree} and \ref{fi_teewith}
as well as our expression for the area spectrum
\eqref{eq_ansatz2} it appears that the pattern is
shifted by $\Delta a/2$ between lines with $2j$
even and lines with $2j$ odd. This explains at
least why the pattern gets washed out
considerably when summed over all $j$. It could
be, however, that a (substantially weaker)
pattern with a spacing of $\Delta a/2$ remains,
and it would be interesting to look for it in
numerical data.

(3) It was observed
\cite{Corichi:2006wn,Corichi:2006bs,Corichi:2007zz,DiazPolo:2007gr}
that $\chi$ is very close to $8\ln(3)$. The
situation here is very tantalizing, in that on
the one hand our result for $\chi$ moves even
closer to the conjectured value. On the other
hand, since our treatment is only approximate, we
can not draw any conclusion from this.

(4) It has been conjectured
\cite{Corichi:2006wn,Corichi:2006bs,Corichi:2007zz,DiazPolo:2007gr}
that the phenomenon continues to be present even
for macroscopic black holes. From our treatment
it does seem that the pattern should start to get
washed out once the black hole is so large that
the dominant paths are longer than about 600
steps. We do not understand the mechanism at work
in the generation of the paths well enough to
present this as a result, however. Rather, we must leave
this question open for future research.


Altogether, we think that the approach taken
affords interesting insights, but it does not
give answers to some crucial questions. Moreover
it may even be questioned wether the present
approach can answer these questions even when
worked out in more detail, since it uses
statistics and some heuristics. Therefor it would
be interesting to pursue alternative approaches.
One possibility that comes to our mind is to
analyze in detail the properties of the
Laplace-Fourier-transform of $n(a,j)$ as
determined to a good approximation in
\cite{Meissner:2004ju} (and which can be
determined \textit{exactly}, as far as we can
see, with similar methods). Its structure, and in
particular its poles, should contain information
on periodic phenomena of $n(a,j)$ along lines
$j=$ \textit{const}. We will pursue this approach
elsewhere.
\section*{Acknowledgements}
I would like to thank A.~Corichi, J.~Diaz-Polo
and E.~Fernandez-Borja for explaining details of
their work to me, for their comments on a draft of this article and
their encouragement, and for interesting
discussions on black hole entropy in loop quantum
gravity. I am grateful to P.~Mitra for pointing
out an error in the bibliography
of an earlier version of this paper, and to
C.~Fleischhack for comments on the genericity
of the phenomenon.

I gratefully acknowledge funding for this work
through a Marie Curie Fellowship of the European
Union.

\end{document}